\documentclass[pdftex,twocolumn,epjc3]{svjour3}          % twocolumn

\RequirePackage[T1]{fontenc}
\RequirePackage{mathptmx}      % use Times fonts if available on your TeX system
\RequirePackage{flushend}
\RequirePackage[numbers,sort&compress]{natbib}
\RequirePackage[colorlinks,citecolor=blue,urlcolor=blue,linkcolor=blue]{hyperref}

\usepackage{bm}
\usepackage{indentfirst}
\usepackage{amsmath}
\usepackage{graphicx}
\usepackage{float}
\usepackage{amssymb}
\usepackage{subfigure}
\usepackage{amssymb}
\usepackage{psfrag}
\usepackage{hyperref}
\usepackage{epstopdf}
\usepackage{diagbox}
\usepackage{CJK}

\journalname{Eur. Phys. J. C}

\begin{document}

\title{Constraints on primordial curvature  spectrum from primordial black holes and  scalar-induced gravitational waves}
\author{  Zhu Yi\thanksref{e1,addr1}
	\and
	Qin Fei\thanksref{e2,addr2}
}

\thankstext{e1}{e-mail: yz@bnu.edu.cn}
\thankstext{e2}{e-mail: feiqin@hbpu.edu.cn}

\institute{Advanced Institute of Natural Sciences, Beijing Normal University, Zhuhai 519087, China \label{addr1} \and 
School of Mathematics and Physics, Hubei Polytechnic University, Huangshi 435003, China \label{addr2}}
\date{Received: date / Accepted: date}

\maketitle
\begin{abstract}
The observational data of primordial black holes and scalar-induced gravitational waves can   constrain  the primordial curvature perturbation at small scales. We parameterize the primordial curvature perturbation by a broken power law form  and find that it is consistent with many inflation models that can produce primordial black holes, such as nonminimal derivative coupling inflation, scalar-tensor inflation, Gauss-Bonnet inflation, and K/G   inflation.  The constraints from primordial black holes on the  primordial curvature power spectrum with the broken power law form  are obtained,
	where the fraction of primordial black holes in dark matter is calculated by the peak theory. 
	Both the real-space top-hat and  the Gaussian window functions are considered. The constraints on the amplitude of primordial curvature perturbation with Gaussian window function are around three times larger than those with  real-space top-hat window function.
	The constraints on the primordial curvature perturbation from the NANOGrav 12.5yrs data sets are  displayed, where the NANOGrav signals are assumed as the scalar-induced gravitational waves, and only the first five frequency bins are used.
\end{abstract}

\section{Introduction}

Primordial black holes (PBHs) are formed from the overdense regions of the Universe by  gravitational collapse  during radiation domination  \cite{Carr:1974nx,Hawking:1971ei}.
They can explain the source of the  gravitational waves (GWs) events detected by the Laser  Interferometer Gravitational Wave Observatory (LIGO)  Scientific  Collaboration and the Virgo Collaboration \cite{Bird:2016dcv,Clesse:2016vqa,Sasaki:2016jop, Abbott:2016blz,Abbott:2016nmj, Abbott:2017vtc, Abbott:2017oio,TheLIGOScientific:2017qsa, Abbott:2017gyy,LIGOScientific:2018mvr,Abbott:2020uma,LIGOScientific:2020stg,  Abbott:2020khf,Abbott:2020tfl,LIGOScientific:2020ibl}.   
They are also dark matter (DM) candidates  \cite{Ivanov:1994pa,Frampton:2010sw,Belotsky:2014kca,Khlopov:2004sc,Clesse:2015wea,Carr:2016drx,Inomata:2017okj,Garcia-Bellido:2017fdg,Kovetz:2017rvv,Carr:2020xqk} due to the non detection of particle dark matter. For the PBHs with masses around $10^{-17}-10^{-15} M_{\odot}$ and $10^{-14}-10^{-12}  M_{\odot}$, they can make up almost all  dark matter because of  no  observational constraints on  the abundances of PBHs at these two mass windows.   
The seed of the overdense regions that collapse to form PBHs can come from the primordial curvature perturbations generated during inflation \cite{Guth:1980zm,Linde:1981mu,Albrecht:1982wi,Starobinsky:1980te}. 
To produce enough PBHs DM, the amplitude of the power spectrum  of the primordial curvature perturbations  should be around $A_\zeta\sim \mathcal{O}(0.01)$. From the observation of  cosmic microwave background (CMB) anisotropy measurements, the  constraints on  the primordial curvature perturbations are $A_\zeta=2.1\times 10^{-9}$ \cite{Akrami:2018odb} at  large scales, $k\lesssim\mathcal{O}(1)\,\text{Mpc}^{-1}$. 
Therefore, to  produce enough PBHs DM, the primordial curvature power spectrum should be   enhanced by  about seven orders of magnitude at small scales,   compared with the constraints at larger scales   \cite{Gong:2017qlj}.  

The amplitude of the primordial curvature perturbations can be sharply enhanced if there is an ultra-slow-roll phase in the inflation model \cite{Martin:2012pe,Motohashi:2014ppa,Yi:2017mxs}.   
For the single field inflation models, the ultra-slow-roll phase can be realized by the canonical inflation models with an inflection point \cite{Garcia-Bellido:2017mdw,Germani:2017bcs,Motohashi:2017kbs,Ezquiaga:2017fvi,Gong:2017qlj,Ballesteros:2018wlw,Dalianis:2018frf,Bezrukov:2017dyv,Passaglia:2018ixg}, and the ultra-slow-roll phase can also be realized by many  noncanonical inflation models \cite{Kamenshchik:2018sig,Fu:2019ttf,Fu:2019vqc,Dalianis:2019vit,Braglia:2020eai,Gundhi:2020zvb,Cheong:2019vzl,Lin:2020goi,Lin:2021vwc,Gao:2020tsa,Gao:2021vxb,Yi:2020kmq,Yi:2020cut,Yi:2021lxc, Yi:2022anu,Zhang:2021rqs,Kawai:2021edk,Cai:2021wzd,Chen:2021nio,Zheng:2021vda}.   
The speed of the enhancement of the primordial curvature perturbations of the single field inflation can not be arbitrarily fast; the steepest enhancement obeys the power law form, $\mathcal{P}_\zeta \sim k^4$ \cite{Byrnes:2018txb,Carrilho:2019oqg}.
Guided by this property, the typical peaks in the power spectrum of the primordial curvature perturbations generated by  single field inflation  can be approximated by a broken power law,  $\mathcal{P}_{ peak}(k)=A(\alpha+\beta)/[\beta(k/k_p)^{-\alpha}+\alpha(k/k_p)^\beta]$  \cite{Vaskonen:2020lbd}.  
Considering the large scales constraints,  the power spectrum of the primordial curvature perturbations can be approximated by 
$\mathcal{P}_{ \zeta}(k)=A(\alpha+\beta)/[\beta(k/k_p)^{-\alpha}+\alpha(k/k_p)^\beta]+A_*(k/k_*)^{n_{s*}-1}$. 
In this paper, we show that the broken power law form of the primordial curvature perturbation is consistent with  many  inflation models such as 
the inflation model with nonminimal derivative coupling \cite{Fu:2019ttf},   scalar-tensor inflation model \cite{Yi:2022anu}, Gauss-Bonnet inflation model \cite{Zhang:2021rqs},  and  inflation model with non-canonical kinetic term (K/G inflation) \cite{Yi:2020cut}.  

The observational data of  PBHs can provide constraints on  the primordial curvature perturbations at small scales. There are many works on this topic. The constraints on the  primordial curvature perturbations from the  PBHs   observational data  with  Press-Schechter theory and Gaussian window function are discussed in ref.  \cite{Sato-Polito:2019hws}.   The constraints considering all  the steps from gravitational collapse  to PBHs formation in real space are obtained in ref. \cite{Kalaja:2019uju}. 
And the constraints   with peak theory are given in ref. \cite{Gow:2020bzo} where the power spectrum of the  primordial curvature  perturbation is chosen as the log-normal form.  
As the broken power law form of the primordial curvature perturbation are consistent with many inflation models, in this paper, 
we use the broken power law form and  give the constraints on 
primordial curvature perturbation from the observational data of PBHs with peak theory. Both Gaussian and real-space top-hat window functions are considered.

Accompanying the formation of PBHs, the large scalar perturbations can induce secondary gravitational waves after the horizon reentry during the radiation-dominated epoch  \cite{Matarrese:1997ay,Mollerach:2003nq,Ananda:2006af,Baumann:2007zm,Garcia-Bellido:2017aan,Saito:2008jc,Saito:2009jt,Bugaev:2009zh,Bugaev:2010bb,Alabidi:2012ex,Orlofsky:2016vbd,Nakama:2016gzw,Inomata:2016rbd,Cheng:2018yyr,Cai:2018dig,Bartolo:2018rku,Bartolo:2018evs,Kohri:2018awv,Espinosa:2018eve,Cai:2019amo,Cai:2019elf,Cai:2019bmk,Cai:2020fnq,Domenech:2019quo,Domenech:2020kqm,Fumagalli:2020adf,Fumagalli:2020nvq,Ashoorioon:2022raz,Pi:2020otn,Yuan:2019fwv,Yuan:2019wwo,Yuan:2019udt,Papanikolaou:2020qtd,Papanikolaou:2021uhe,Papanikolaou:2022hkg}.   
The scalar-induced gravitational waves (SIGWs), containing much information about the early Universe,   have wide frequency distribution and can be detected by  pulsar timing arrays (PTA) \cite{Ferdman:2010xq,Hobbs:2009yy,McLaughlin:2013ira,Hobbs:2013aka,Moore:2014lga} and
the future space-based GW detectors such as Laser Interferometer Space Antenna (LISA) \cite{Danzmann:1997hm,Audley:2017drz}, Taiji \cite{Hu:2017mde}, and TianQin  \cite{Luo:2015ght}.  
The stochastic process  with a common amplitude and a common spectral slope across  pulsars  detected by 
the North American Nanohertz Observatory for Gravitational Wave (NANOGrav)  Collaboration \cite{NANOGrav:2020bcs} and other pulsar timing arrays \cite{Goncharov:2021oub,Antoniadis:2022pcn} can be explained by SIGWs \cite{DeLuca:2020agl,Inomata:2020xad, Vaskonen:2020lbd, Domenech:2020ers, Yi:2021lxc}.  In this paper, we give the constraints on the primordial curvature perturbation from the NANOGrav 12.5yrs data sets  by regarding the NANOGrav signals as GWs  induced  from the primordial curvature perturbation with the broken power law form.

This paper is organized as follows. In Sec. 2, we calculate the abundance of PBHs  from the power spectrum of primordial curvature perturbation by  peak theory.  In   Sec. 3,  we give  the  energy density of SIGWs.  
We discuss the constraints from the PBHs and SIGWs on the power spectrum of primordial curvature perturbation in Sec. 4.
We conclude the paper in Sec. 5.

\section{The primordial black holes}
The fraction of the Universe in PBHs at the formation is denoted by
\begin{equation}\label{beta}
	\beta=\frac{\rho_{\text{PBH}}}{\rho_b},
\end{equation}
where $\rho_\text{PBH}$ is the energy density of  PBHs and $\rho_b$ is the background energy density of the Universe. 
From the peak theory, the energy density of PBHs  is 
\cite{Bardeen:1985tr,Green:2004wb,Young:2014ana,Germani:2018jgr,Young:2020xmk,Gow:2020bzo}
\begin{equation}\label{rho:pbh} 
	\rho_{\text{PBH}}=\int_{\nu_c}^{\infty}M_{\text{PBH}}(\nu)\mathcal{N}_{pk}(\nu)d\nu,
\end{equation}
where $\nu=\delta/\sigma$ and $\nu_c=\delta_c/\sigma_0$,  $\delta_c$ is the threshold of  the  smoothed density contrast  for the formation of PBHs and  $\sigma_0$ is the variance of the smoothed density contrast.   $M_\text{PBH}$ is the mass of  PBHs  and $	\mathcal{N}_{pk}$ is  the number density of PBHs \cite{Bardeen:1985tr},
\begin{equation}\label{num:den}
	\mathcal{N}_{pk}(\nu)=\frac{1}{a^3}\frac{1}{(2\pi)^2}\left(\frac{\sigma_1}{\sqrt{3}\sigma_0}\right)^3
	\nu^3\exp\left(-\frac{\nu^2}{2}\right).
\end{equation}
The moment of the smoothed density power spectrum $\sigma_n$  is  defined by
\begin{equation}\label{variance1}
	\sigma^2_n=\int_{0}^{\infty}\frac{dk}{k}k^{2n} T^2(k,R_H)W^2(k,R_H)\mathcal{P}_\delta(k),
\end{equation}
where $\sigma_0$ and $\sigma_1$ are obtained by choosing $n=0$ and $n=1$, respectively.
The   power spectrum of the density contrast   $\mathcal{P}_\delta$  is related to the 
power spectrum of  primordial curvature  perturbations  $\mathcal{P}_\zeta$   by
\begin{equation}\label{rel:pp}
	\mathcal{P}_\delta(k)=\frac{4(1+w)^2}{(5+3w)^2}\left(\frac{k}{aH}\right)^4 \mathcal{P}_{\zeta}(k),
\end{equation}
where the state equation $w=1/3$ during the radiation domination. The most considered   window functions $W(k,R_H)$ in equation \eqref{variance1} are  the real-space top-hat window function \cite{Ando:2018qdb},
\begin{equation}\label{window:fun}
	W(k,R_H)=3\left[\frac{\sin\left(kR_H\right)-\left(kR_H\right)
		\cos\left(kR_H\right)}{\left(kR_H\right)^3}\right],
\end{equation}
and the  Gaussian window function
\begin{equation}\label{window:fun:G}
	W(k,R_H)=\exp\left(\frac{-k ^2 R_H^2}{2}\right),
\end{equation}
with the  smoothed scale    $R_H\sim 1/aH$.  In this paper,  both these two window functions are considered.
The transfer function in equation \eqref{variance1} is
\begin{equation}\label{transfer}
	T(k,R_H)=3\left[\frac{\sin\left(\frac{kR_H}{\sqrt{3}}\right)-\left(\frac{kR_H}{\sqrt{3}}\right)
		\cos\left(\frac{kR_H}{\sqrt{3}}\right)}{\left({kR_H}/{\sqrt{3}}\right)^3}\right].
\end{equation}

The mass  of primordial black holes in equation \eqref{rho:pbh}  obeys the   critical scaling law \cite{Choptuik:1992jv,Evans:1994pj,Niemeyer:1997mt},
\begin{equation}\label{pbh:mass}
	M_{\text{PBH}}=\kappa M_H(\delta-\delta_c)^{\gamma},
\end{equation}
where $\gamma=0.36$ in the radiation domination \cite{Choptuik:1992jv,Evans:1994pj} and $M_H$ is the mass  in the horizon,
\begin{equation}\label{mass:h}
	M_H\approx 13\left(\frac{g_*}{106.75}\right)^{-1/6}\left(\frac{k}{10^6 \text{Mpc}^{-1}}\right)^{-2}M_\odot,
\end{equation}
$g_*$ is the number of relativistic degrees of freedom at the formation of PBHs. The parameter $\kappa$ in the critical scaling law and  PBHs formation threshold  $\delta_c$  are  dependent on the window function, and the relations of them between the two window functions are  \cite{Young:2020xmk}
\begin{gather}\label{two:windows}
	\delta_{c(G)}\approx\frac{\delta_{c(TH)}}{2.17},\quad \kappa_G\approx\frac{2.74^2\times 2.17^\gamma}{4} 	\kappa_{TH},
\end{gather}
where $\gamma$ is the index in the  critical scaling law \eqref{pbh:mass}. 
For the real-space top-hat window function,  we choose  $\delta_{c(TH)}=0.51$  and  $\kappa_{TH}=3.3$ \cite{Young:2019osy,Musco:2018rwt}; from equation \eqref{two:windows},  the corresponding values for the Gaussian window function  are $\delta_{c(G)}\approx 0.24$  and  $\kappa_{G}=8.2$.

Combining the mass fraction of the Universe which   collapses to form PBHs, equation \eqref{beta}, with the background equation of the energy density of the Universe,  we obtain the density parameter of  PBHs  at present  \cite{Byrnes:2018clq},
\begin{equation}\label{beta:omega}
	\Omega_\text{PBH}=\int_{M_\text{min}}^{M_\text{max}} d \ln M_H \left(\frac{M_{eq}}{M_H}\right)^{1/2}\beta(M_H),
\end{equation}
where we use the relations $\rho_\text{PBH}\propto a^{-3}$  and  $\rho_b\propto a^{-n}$, with $n=4$ during radiation domination and $n=3$ during  matter domination; and $M_{eq}=2.8\times 10^{17}M_{\odot}$ is the horizon mass at the matter-radiation equality.   The limits of the integral are taken as $M_\text{min}=0$ and $M_\text{max}=\infty$ because of $\beta(M_H)\rightarrow 0$ at the condition $M_H\rightarrow 0$ or $M_H\rightarrow\infty$. The fraction of primordial black holes in the dark matter at present  is defined as
\begin{equation}\label{fpbh:tot}
	f_{\text{PBH}}=\frac{\Omega_{\text{PBH}}}{\Omega_\text{DM}}=\int f(M_\text{PBH}) d\ln M_\text{PBH},
\end{equation}
where $\Omega_\text{DM}$ is the density parameter for  dark matter, and the PBHs mass function is
\begin{equation}\label{mass:func}
	f(M_\text{PBH})=\frac{1}{\Omega_{\text{DM}}}\frac{d \Omega_{\text{PBH}}}{d \ln M_{\text{PBH}}}.
\end{equation}
Substituting  equation \eqref{beta:omega} into definition  \eqref{mass:func} and combining the above equations,  
we can obtain the PBHs mass function  \cite{Byrnes:2018clq},
\begin{equation}\label{app:fpbh:beta}
	\begin{split}
		f(M_\text{PBH})&=\frac{1}{\Omega_{\text{DM}}} \int_{M_\text{min}}^{M_\text{max}}\frac{d M_H}{M_H}
		\frac{M_\text{PBH}}{\gamma M_H} \sqrt{\frac{M_{eq}}{M_H}}\\
		&\times\frac{1}{3\pi} \left(\frac{\sigma_1}{\sqrt{3}\sigma_0 aH}\right)^3\frac{1}{\sigma_0^4}
		\left(\mu^{1/\gamma}+\delta_c\right)^3\\
		&\times \mu^{1/\gamma} \exp\left[-\frac{\left(\mu^{1/\gamma}+\delta_c\right)^2}{2\sigma_0^2}\right],
	\end{split}
\end{equation}
where $\mu=M_{\text{PBH}}/(\kappa M_H)$ and the  relation  $d\delta/d\ln M_{\text{PBH}}=\mu^{1/\gamma}/\gamma$ derived from equation   \eqref{pbh:mass} is used.

\section{The scalar-induced gravitational waves}
Accompanying  the formation of PBHs, the large scalar perturbation can induce secondary gravitational waves during the radiation domination.
The metric with perturbation in the cosmological background  and  Newtonian gauge is
\begin{equation}
	\begin{split}
		d s^2=&-a^2(\eta)(1+2\Phi)d\eta^2 \\
		&+a^2(\eta)\left[(1-2\Phi)\delta_{ij}+\frac12h_{ij}\right]d x^i d x^j,
	\end{split}
\end{equation}
where the anisotropic stress is  neglected, $\eta$ is the conformal time, $\Phi$ is the Bardeen potential,  
and $h_{ij} $ are the  tensor perturbations. The tensor perturbations in the Fourier space can be obtained by the transform
\begin{equation}
	\label{hijkeq1}
	h_{ij}(\bm{x},\eta)=\int\frac{  d^3k  e^{i\bm{k}\cdot\bm{x}}}{(2\pi)^{3/2}}
	[h_{\bm{k}}(\eta)e_{ij}(\bm{k})+\tilde{h}_{\bm{k}}(\eta)\tilde{e}_{ij}(\bm{k})],
\end{equation}
where $e_{ij}(\bm{k})$ and $\tilde{e}_{ij}(\bm{k})$  are the plus and cross polarization tensors,
\begin{gather}
	e_{ij}(\bm{k})=\frac{1}{\sqrt{2}}\left[e_i(\bm{k})e_j(\bm{k})-\tilde{e}_i(\bm{k})\tilde{e}_j(\bm{k})\right], \\
	\tilde{e}_{ij}(\bm{k})=\frac{1}{\sqrt{2}}\left[e_i(\bm{k})\tilde{e}_j(\bm{k})+\tilde{e}_i(\bm{k})e_j(\bm{k})\right],
\end{gather}
with the basis vectors satisfying $\bm e\cdot \tilde{\bm e}=\bm e \cdot \bm{k}= \tilde{\bm e}\cdot\bm{k}$.

The tensor perturbations  in the Fourier space with either  polarization induced by the second order of the linear scalar perturbations satisfy \cite{Ananda:2006af,Baumann:2007zm}
\begin{equation}
	\label{eq:hk}
	h''_{\bm{k}}+2\mathcal{H}h'_{\bm{k}}+k^2h_{\bm{k}}=4S_{\bm{k}},
\end{equation}
where a prime denotes the derivative with respect to the conformal time $\eta$,
$h'_{\bm{k}}=dh_{\bm{k}}/d\eta$, and $\mathcal{H}=a'/a $ is the conformal Hubble parameter. The second order source from the linear scalar perturbations $S_{\bm{k}}$ is 
\begin{equation}
	\label{hksource}
	\begin{split}
		S_{\bm{k}}=&\int \frac{d^3\tilde{k}}{(2\pi)^{3/2}}e_{ij}(\bm{k})\tilde{k}^i\tilde{k}^j
		\left[2\Phi_{\tilde{\bm{k}}}\Phi_{\bm{k}-\tilde{\bm{k}}} \phantom{\frac{1}{2}}+ \right.\\
		&\left.\frac{1}{\mathcal{H}^2} \left(\Phi'_{\tilde{\bm{k}}}+\mathcal{H}\Phi_{\tilde{\bm{k}}}\right)
		\left(\Phi'_{\bm{k}-\tilde{\bm{k}}}+\mathcal{H}\Phi_{\bm{k}-\tilde{\bm{k}}}\right)\right],
	\end{split}
\end{equation}
where  $\Phi_{\bm{k}}$  is the  Bardeen potential in the Fourier space and can be related to the primordial curvature  perturbations $\zeta_{\bm{k}}$ generated in the inflation by the transfer function 
\begin{equation}
	\Phi_{\bm{k}}=\frac{3+3w}{5+3w}T(k,\eta) \zeta_{\bm{k}},
\end{equation}
where  $T(k,\eta)$ is  the transfer function \eqref{transfer}.  

To solve the  tensor perturbations  equation \eqref{eq:hk}, the  Green function method is used, and the solution is 
\begin{equation}\label{hk:green}
	h_k(\eta)=\frac{4}{a(\eta)}\int_{\eta_k}^{\eta}d \tilde{\eta}g_k(\eta,\tilde{\eta})a(\tilde{\eta})S_k(\tilde{\eta}),
\end{equation}
where the   corresponding  Green function is
\begin{equation}\label{green}
	g_k(\eta,\eta')=\frac{\sin\left[k(\eta-\eta')\right]}{k}.
\end{equation}
Substituting the solution of  $h_k$ \eqref{hk:green}  into the definition of the power spectrum of  tensor perturbations,
\begin{equation}
	\label{eq:pwrh}
	\langle h_{\bm{k}}(\eta)h_{\tilde{\bm{k}}}(\eta)\rangle
	=\frac{2\pi^2}{k^3}\delta^{(3)}(\bm{k}+\tilde{\bm{k}})\mathcal{P}_h(k,\eta),
\end{equation}
we can    obtain  \cite{Baumann:2007zm,Ananda:2006af,Kohri:2018awv,Espinosa:2018eve,Lu:2019sti}
\begin{equation}\label{ph}
	\begin{split}
		\mathcal{P}_h(k,\eta)=&
		4\int_{0}^{\infty}dv\int_{|1-v|}^{1+v}du \left[\frac{4v^2-(1-u^2+v^2)^2}{4uv}\right]^2\\ &\times I_{RD}^2(u,v,x)\mathcal{P}_{\zeta}(kv)\mathcal{P}_{\zeta}(ku),
	\end{split}
\end{equation}
where $u=|\bm{k}-\tilde{\bm{k}}|/k$, $v=\tilde{k}/k$, $x=k\eta$, and the integral kernel $I_{\text{RD}}$  is
\begin{equation}
	\label{irdeq1}
	\begin{split}
		I_{\text{RD}}(u, v, x)=&\int_1^x dy\, y \sin(x-y)\{3T(uy)T(vy)\\
		&+y[T(vy)u T'(uy)+v T'(vy) T(uy)]\\
		&+y^2 u v T'(uy) T'(vy)\}.
	\end{split}
\end{equation}
The energy density of Gravitational waves is defined as 
\begin{equation}
	\label{density}
	\Omega_{\mathrm{GW}}(k,\eta)=\frac{1}{24}\left(\frac{k}{aH}\right)^2\overline{\mathcal{P}_h(k,\eta)}.
\end{equation}
Substituting equation \eqref{ph} into the definition  \eqref{density},  we get \cite{Espinosa:2018eve,Lu:2019sti}
\begin{equation}
	\label{SIGWs:gwres1}
	\begin{split}
		\Omega_{\mathrm{GW}}(k,\eta)=&\frac{1}{6}\left(\frac{k}{aH}\right)^2\int_{0}^{\infty}dv\int_{|1-v|}^{1+v}du \\
		&\times\left[\frac{4v^2-(1-u^2+v^2)^2}{4uv}\right]^2\\
		&\times\overline{I_{\text{RD}}^{2}(u, v, x)} \mathcal{P}_{\zeta}(kv)\mathcal{P}_{\zeta}(ku),
	\end{split}
\end{equation}
where $\overline{I_{\text{RD}}^{2}}$ is the oscillation time average of the integral kernel. 
The evolution of the energy  density of the gravitational waves is the same as that of radiation; with the help of  this property,  the energy density of  the gravitational waves at present can be obtained easily,   
\begin{equation}\label{d}
	\Omega_{\mathrm{GW}}(k,\eta_0)=\frac{c_g\Omega_{r,0}  \Omega_{\mathrm{GW}}(k,\eta)}{\Omega_{r}(\eta)},
\end{equation}
where  $\Omega_{r,0}$ is the energy density of the radiation at present and $\Omega_{r}(\eta)=1$ at the generation of SIGWs,  and   \cite{Vaskonen:2020lbd,DeLuca:2020agl}
\begin{equation}\label{gwcg}
	c_g=0.387\left(\frac{g_{*,s}^4g_*^{-3}}{106.75}\right)^{-1/3}.
\end{equation}

\section{The constraints on the primordial power spectrum}
At large scales,  the constraints on the primordial curvature perturbations from the observation of  CMB anisotropy measurements are strong,  $A_\zeta=2.1\times 10^{-9}$ \cite{Akrami:2018odb}. At the same time, there are few constraints on the primordial curvature perturbations at small scales.  
Because  PBHs and scalar-induced GWs are produced from  the primordial curvature perturbations with large amplitude at small scales,  the successful or failed detection of PBHs and SIGWs can provide  constraints on  the primordial curvature perturbations at small scales.   
To produce  enough  PBHs DM and  SIGWs,   the power spectrum of the primordial perturbations should be around $\mathcal{P}_\zeta \sim \mathcal{O}(0.01)$, which is  about seven orders of magnitudes larger than that at large scales. 
For the single field inflation models, the profile of  the enhancement of the power spectrum can be governed by the power law 
form, and the  steepest enhancement is about the order of $\sim k^4$ \cite{Byrnes:2018txb,Carrilho:2019oqg}. To fit the enhanced primordial power spectrum, we consider the broken power law  parameterization  \cite{Vaskonen:2020lbd}
\begin{equation}\label{bpl:parametrize}
	\mathcal{P}_\zeta(k)=\frac{A(\alpha+\beta)}{\beta(k/k_p)^{-\alpha}+\alpha(k/k_p)^\beta}+A_*(k/k_*)^{n_{s_*}-1}.
\end{equation}
The first term is the broken power law form to fit the enhanced peak at small scales, $0.5 \lesssim   \alpha \lesssim4$  controls the speed of enhancement, and  $0.5 \lesssim \beta  \lesssim4$ determines the speed of  decline in the power spectrum. A pair of smaller parameters, $\alpha$ and $\beta$,   gives a broader peak in the spectrum. The lower limit of the parameters  ensures the curvature power spectrum between the end of inflation and the peak obeys  a power law \cite{Vaskonen:2020lbd}.
The  second term in equation \eqref{bpl:parametrize} is the near scale-invariant power law form to fit the Planck 2018 constraints  \cite{Planck:2018jri} at large scales, and the parameters are chosen as
$k_*=0.05\text{Mpc}^{-1}$, $n_{s_*}=0.965$, and $A_*=2.1\times 10^{-9}$.   

\begin{figure*}[htbp]
	\centering
	\includegraphics[width=0.95\columnwidth]{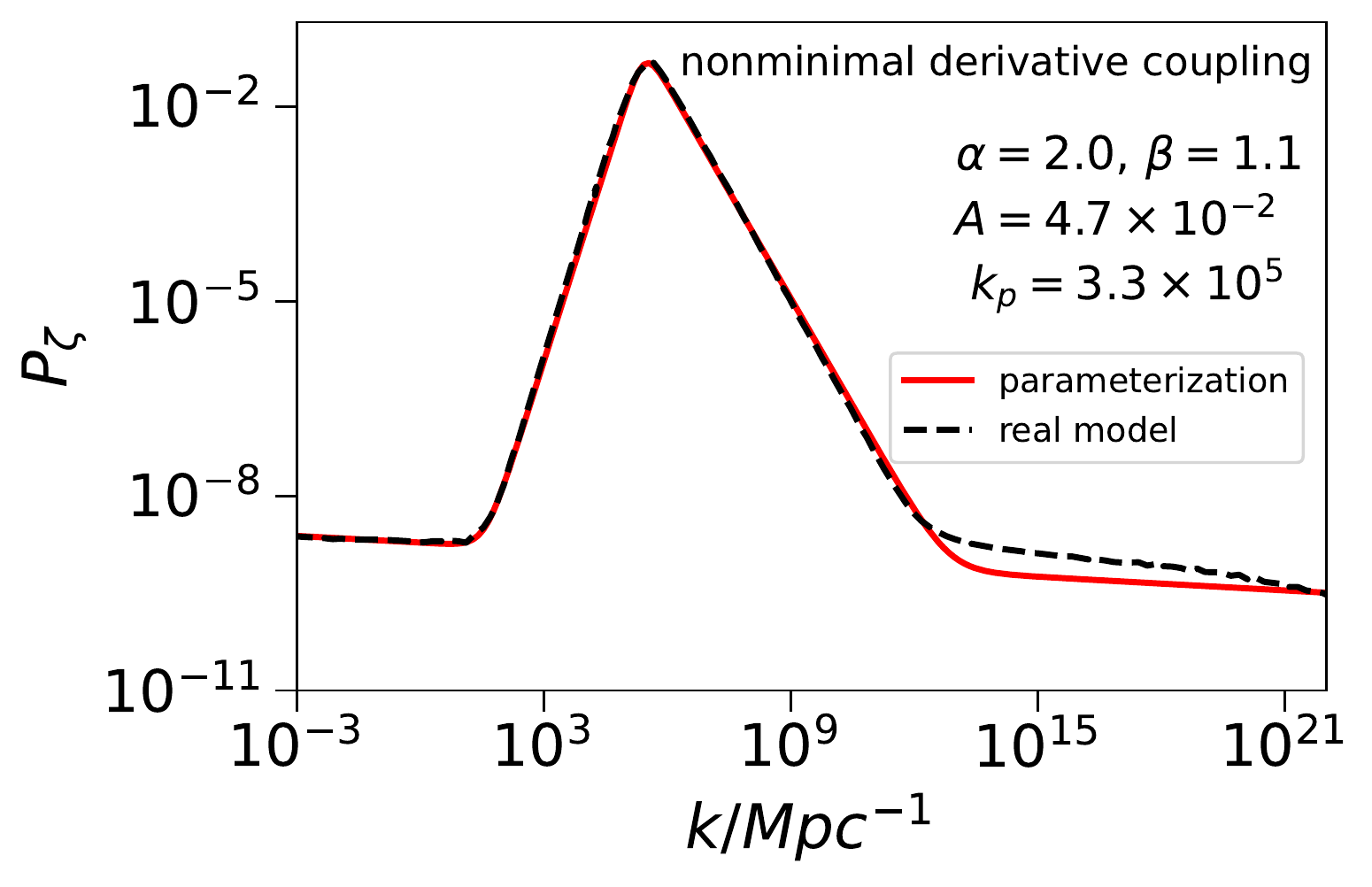}
	\includegraphics[width=0.95\columnwidth]{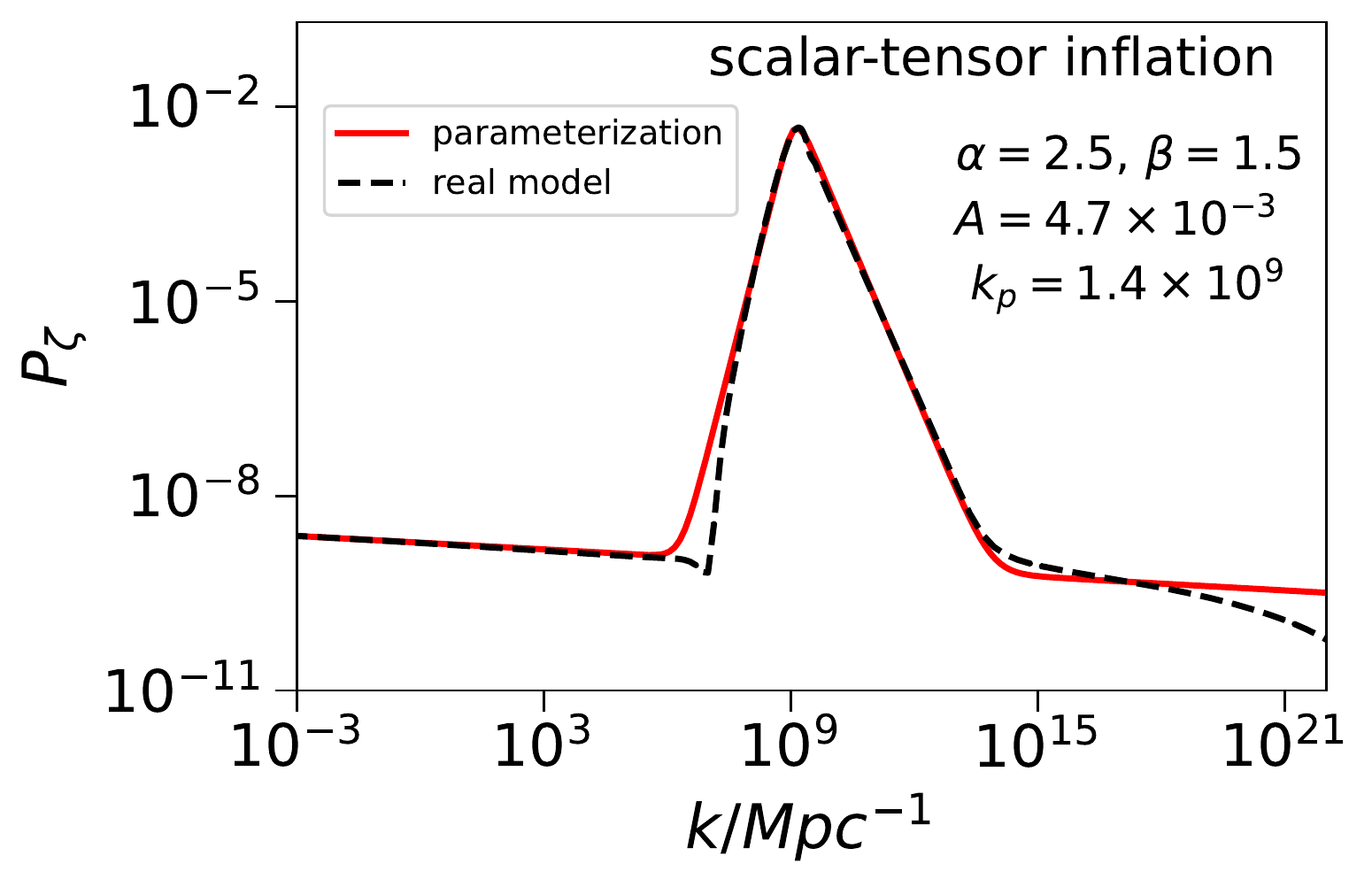}
	\includegraphics[width=0.95\columnwidth]{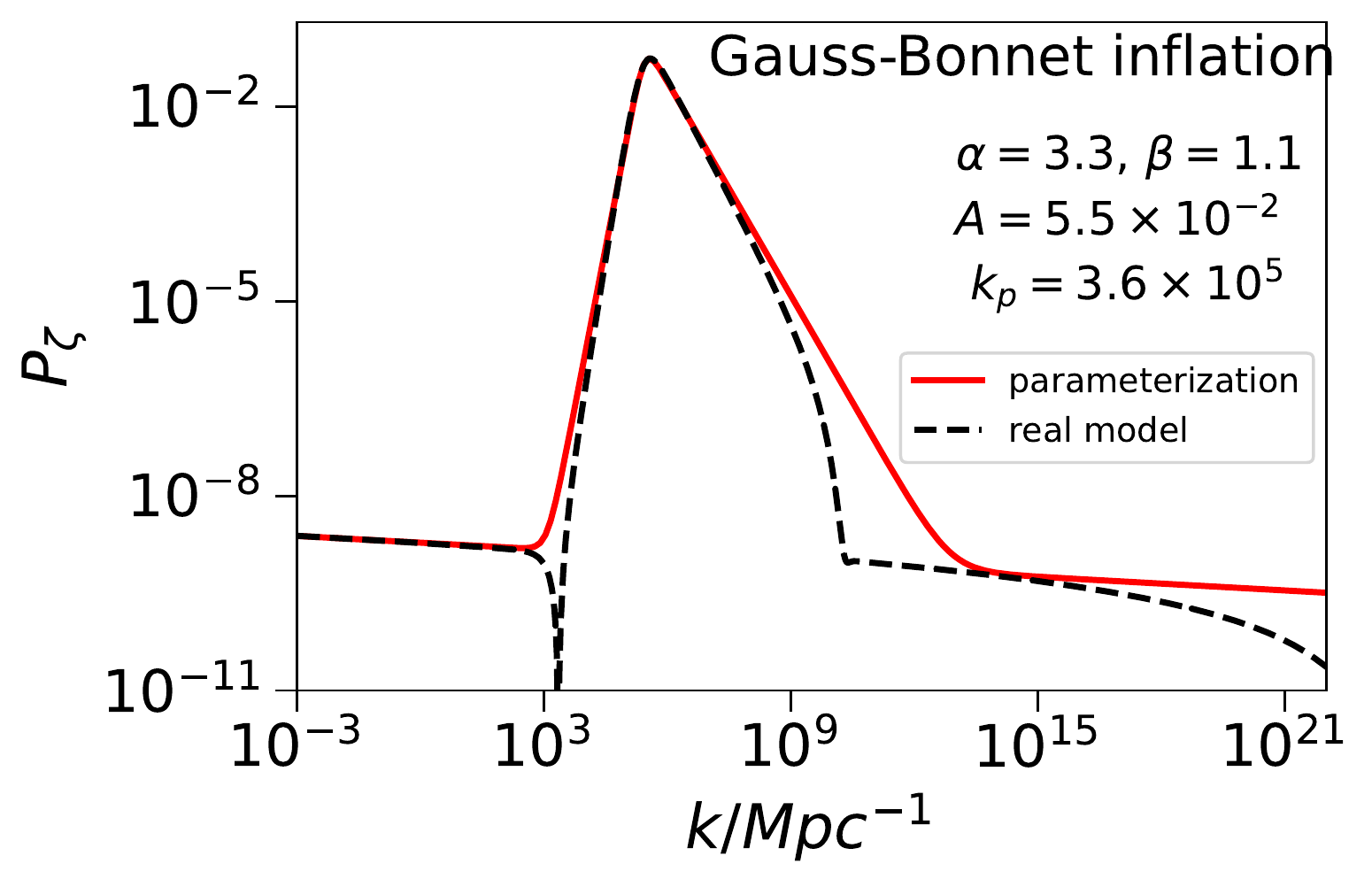}
	\includegraphics[width=0.95\columnwidth]{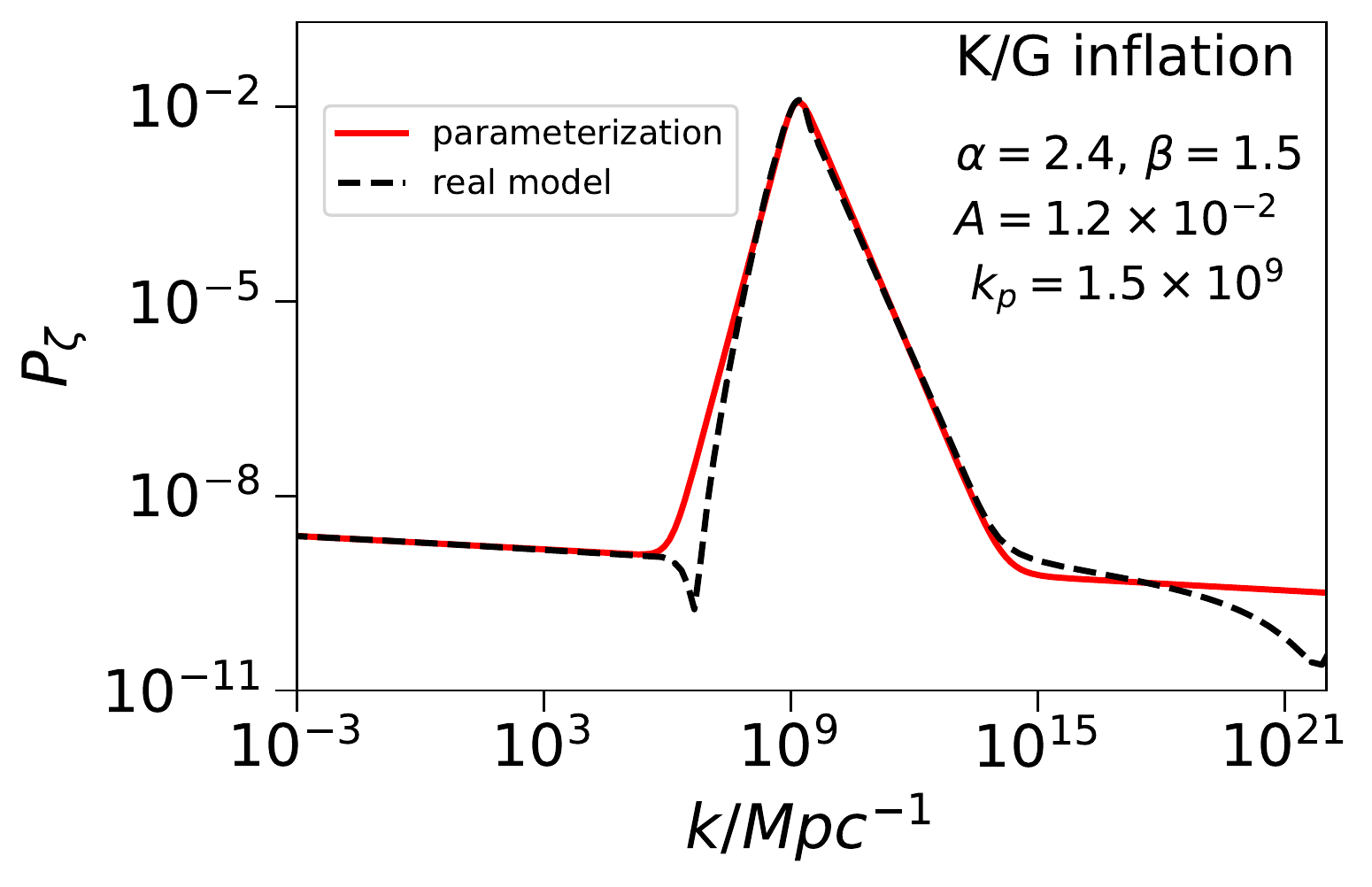}
	\caption{The difference between the power spectra from real inflation models (denoted by black dashed lines) and parameterization \eqref{bpl:parametrize} (marked by solid red lines). 
		The left upper panel, right upper panel, left lower panel, and right lower panel are the power spectra from inflation  model with nonminimal derivative coupling \cite{Fu:2019ttf},   scalar-tensor inflation model \cite{Yi:2022anu}, Gauss-Bonnet inflation model \cite{Zhang:2021rqs},  and  inflation model with non-canonical kinetic term (K/G inflation) \cite{Yi:2020cut}, respectively.}\label{fig:compare} 
\end{figure*}

In figure \ref{fig:compare}, we compare  the parameterization \eqref{bpl:parametrize}  with some real inflation models.  The black dashed lines  denote the  power spectra from real inflation models, and the red solid lines represent the results from   parameterization \eqref{bpl:parametrize}. 
The  power spectra in the left upper panel, right upper panel, left lower panel, and right lower panel are from  the  inflation  model with nonminimal derivative coupling \cite{Fu:2019ttf},   scalar-tensor inflation model \cite{Yi:2022anu}, Gauss-Bonnet inflation model \cite{Zhang:2021rqs}, and inflation model with a non-canonical kinetic term (K/G inflation model) \cite{Yi:2020cut}, respectively. 
Figure \ref{fig:compare} shows that the power spectra from parameterization \eqref{bpl:parametrize} are consistent well with those from real inflation models both at   large and peak scales. The features of PBHs and SIGWs are mainly determined by the high peak region of  the power spectrum of the primordial curvature perturbations. Therefore, we can safely use  parameterization \eqref{bpl:parametrize} to research the topic of PBHs and SIGWs. 

\subsection{Constraints from PBHs}
\begin{figure*}[htbp]
	\centering
	\includegraphics[width=0.95\columnwidth]{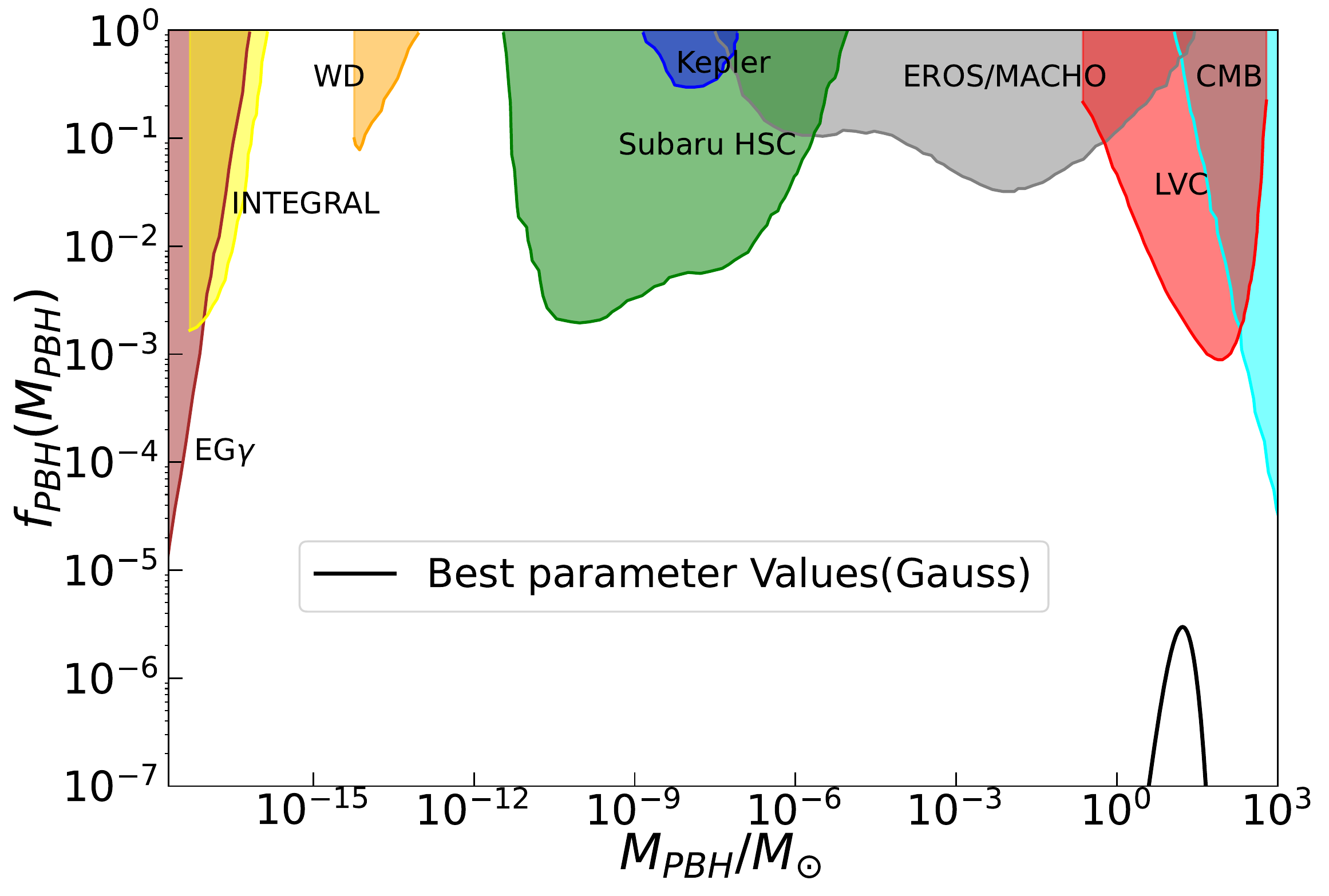}
	\includegraphics[width=0.95\columnwidth]{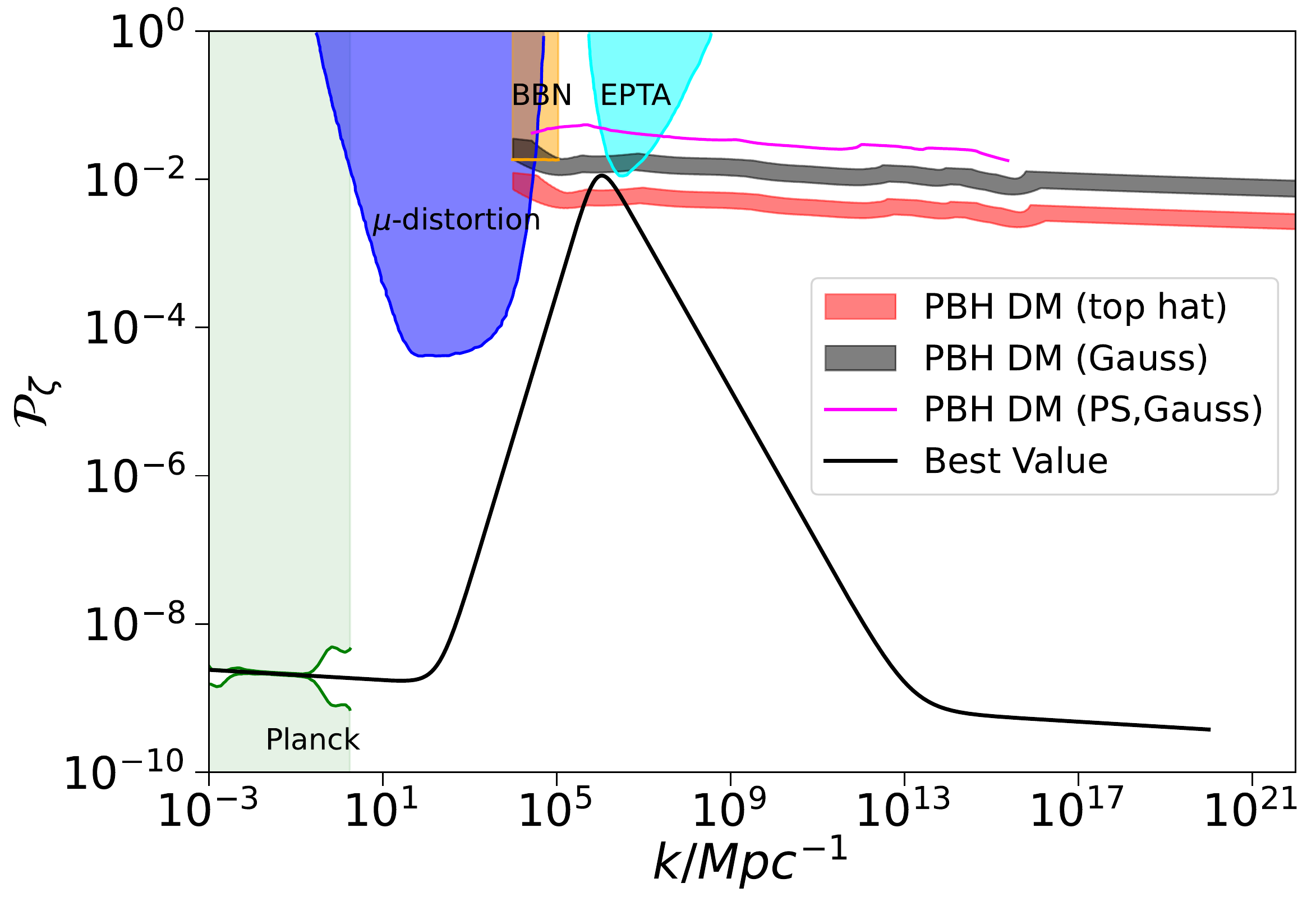}
	\caption{The left panel shows the observational constraints on the PBHs abundance:
		the cyan region from accretion constraints by CMB \cite{Ali-Haimoud:2016mbv,Poulin:2017bwe},
		the red region from LIGO-Virgo Collaboration measurements \cite{Ali-Haimoud:2017rtz,Raidal:2018bbj,Vaskonen:2019jpv,DeLuca:2020qqa,Wong:2020yig,Hutsi:2020sol},
		the gray region from the EROS/MACHO \cite{Tisserand:2006zx},
		the green region from microlensing events with Subaru HSC \cite{Niikura:2017zjd},
		the blue region from the Kepler satellite \cite{Griest:2013esa}, the orange region from white dwarf explosion (WD) \cite{Graham:2015apa}, the yellow region from galactic center 511 keV gamma-ray line (INTEGRAL) \cite{Laha:2019ssq,Dasgupta:2019cae,Laha:2020ivk},  and
		the brown region from extragalactic gamma-rays by PBH evaporation (EG$\gamma$) \cite{Carr:2009jm}. 
		The right panel shows the constraints on the primordial curvature spectrum. 
		The light green shaded region is excluded by the CMB observations \cite{Planck:2018jri}.
		The cyan,  orange and blue regions show the constraints from the EPTA observations \cite{Inomata:2018epa},
		the effect on the ratio between neutron and proton during the big bang nucleosynthesis (BBN) \cite{Inomata:2016uip}
		and $\mu$-distortion of CMB \cite{Fixsen:1996nj}, respectively. 
		The red and black bands are from the left panel PBHs observational data.  The black curve in the right panel is the  parameterization \eqref{bpl:parametrize} with the best parameter values \eqref{beta:value1} and \eqref{beta:value2} obtained from the Bayesian analysis of PTA data, and the black curve in the left panel is 
	   the corresponding mass function calculated by the peak theory with the Gaussian  window function.} 
	\label{fig:pl_allpbh_con}
\end{figure*}

Combining  the parameterization \eqref{bpl:parametrize} and  PBHs mass function \eqref{app:fpbh:beta}, 
we can transfer the constraints on  PBHs DM to  those on the primordial curvature power spectrum.  
The left panel of figure \ref{fig:pl_allpbh_con} shows the main   observational data of PBHs DM at present, and the right panel of figure \ref{fig:pl_allpbh_con} displays the corresponding constraints on the primordial curvature power spectrum from the left panel PBHs DM observational data.
In  the right panel of figure \ref{fig:pl_allpbh_con}, the  red  and black bands are the results with the choice of top-hat and Gaussian window functions, respectively.  The lower and upper limits of each band are from the  parameterization \eqref{bpl:parametrize} with $\alpha=\beta=0.5$ and $\alpha=\beta=4$, respectively.  
For the choice of the top-hat window function, the constraints on the primordial curvature power spectrum from PBHs DM are  more robust than those from BBN and PTA; and for the choice  of the Gaussian window function, the constraints are weaker than those from   PTA.
The uncertainty of the constraints caused by the choice of the window function is more significant than that by the profile of the peak in the primordial curvature power spectrum. 
The magenta line denotes the constraints from PBHs DM with  Press-Schechter theory and  Gaussian window function given in ref. \cite{Sato-Polito:2019hws},  
and the constraints are weaker than that with peak theory. The reason is that the PBHs abundance calculated by the peak theory is larger than that by the Press-Schechter by about one order magnitude for the same primordial power spectrum \cite{Green:2004wb}.

\begin{figure}[htbp]
	\centering
	\includegraphics[width=0.95\columnwidth]{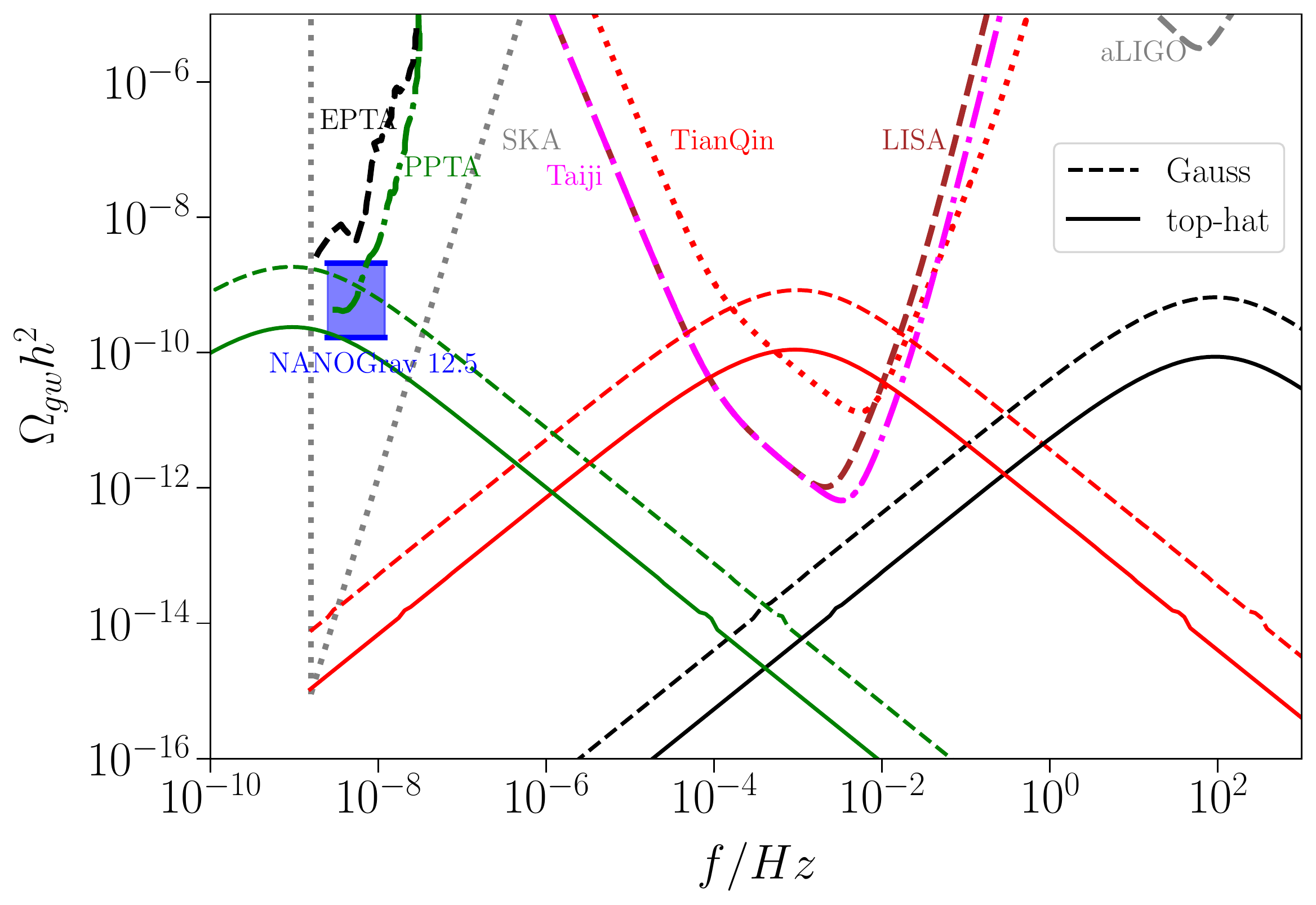}
	\caption{The  convex curves denote the energy density of  SIGWs  with peak frequencies around $10^{-9}$ Hz, $10^{-3}$ Hz, and $10^{2}$ Hz, respectively. 
		The dashed and solid lines are related to  the black and red bands in the right panel of figure \ref{fig:pl_allpbh_con}, respectively.		
		The concave curves represent the GWs detectors limits: 
		the black dashed curve denotes the EPTA limit \cite{Ferdman:2010xq,Hobbs:2009yy,McLaughlin:2013ira,Hobbs:2013aka,Lentati:2015qwp},
		the green dot-dashed curve denotes the PPTA limit \cite{Shannon:2015ect},
		the gray dotted curve denotes the SKA limit \cite{Moore:2014lga},
		the red  dotted curve in the middle denotes the TianQin limit \cite{Luo:2015ght},
		the magenta dot-dashed curve shows the Taiji limit \cite{Hu:2017mde},
		the brown dashed curve shows the LISA limit \cite{Audley:2017drz}, and the gray dashed curve in the right denotes the aLIGO limit \cite{Harry:2010zz,TheLIGOScientific:2014jea}.}
	\label{fig:pl_allpbh_con:gw}
\end{figure}

Accompanying  the PBHs formation, the large scalar perturbations can induce the secondary gravitational waves.  
Substituting the parameterization \eqref{bpl:parametrize} into equation \eqref{d}, we can obtain the energy density of the corresponding SIGWs. 
According to the  lower limits of each band displayed in the right panel of figure \ref{fig:pl_allpbh_con}, we choose the parameter values listed in table \ref{tab:sigw} 
to generate  SIGWs  with  peak frequencies around $10^{-9}$ Hz, $10^{-3}$ Hz, and $10^2$ Hz, and the results  are shown  in figure \ref{fig:pl_allpbh_con:gw}. 
The labels ``Gauss" and ``TopHat"  in table \ref{tab:sigw} denote the choices according to the lower limits of the black and red bands in  the right panel of figure \ref{fig:pl_allpbh_con}, respectively. The parameter $f_p$ is the peak frequency of the SIGW. 
\begin{table}[htbp]
	%\begin{table}[htp]
	\renewcommand\tabcolsep{5.0pt}
	\centering
	\begin{tabular}{llll}
		\hline
		Model \quad   &A& $k_p$& $f_p/\text{Hz}$\\
		\hline
		Gauss1 \quad   &$1.23\times 10^{-2}$&$6.36\times 10^5$&$ 9.83\times 10^{-10}$\\
		Gauss2 \quad   &$8.30\times 10^{-3}$&$6.72\times 10^{11}$&$ 1.04\times 10^{-3}$\\
		Gauss3 \quad  &$7.34\times 10^{-3}$&$6.39\times 10^{16}$&$ 9.88\times 10^{1}$\\
		\hline
		TopHat1 \quad   &$4.42\times 10^{-3}$&$6.51\times 10^5$&$ 1.01\times 10^{-9}$\\
		TopHat2 \quad   &$3.02\times 10^{-3}$&$6.37\times 10^{11}$&$ 9.84\times 10^{-4}$\\
		TopHat3 \quad   &$2.67\times 10^{-3}$&$6.35\times10^{16} $&$9.82\times10^{1}$\\
		\hline
	\end{tabular}
	\caption{The parameter values of the parameterization \eqref{bpl:parametrize}  with  $\alpha=\beta=0.5$. }
	\label{tab:sigw}
\end{table} 
The solid  and  dashed lines in figure \ref{fig:pl_allpbh_con:gw} are the energy density of the SIGWs with the parameter  value choices labeled  as "TopHat" and "Gauss" in table \ref{tab:sigw}, respectively.  The green, red, and black lines  
denote the SIGWs with peak frequencies around  $10^{-9}$ Hz, $10^{-3}$ Hz, and $10^2$ Hz, respectively.  
The SIGWs with $10^{-9}$ Hz  can explain the  stochastic  
common-spectrum process detected by the NANOGrav and other PTA  groups recently. In the future, more and more SIGWs with peak frequency around $10^{-9}$ Hz may be   detected by  PTA groups.
The  SIGWs with $10^{-3}$ Hz can be detected by the future space-based GWs detectors such as LISA, Taiji, and TianQin. 
Under the present PBHs DM observational data constraints,  the corresponding SIGWs can be detected by both the present PTA and future space-based GW detectors. More information about inflation at small scales will be obtained with more observational data about SIGWs being detected. 
The SIGWs with  peak frequencies around $10^2$ Hz cannot be  detected  by the  aLIGO detector; therefore, the aLIGO detector can tell us little information about the inflation at small scales  through  SIGWs at present.

\subsubsection{Constraints from all dark matter}
\begin{figure}[htbp]
	\centering
	\includegraphics[width=0.95\columnwidth]{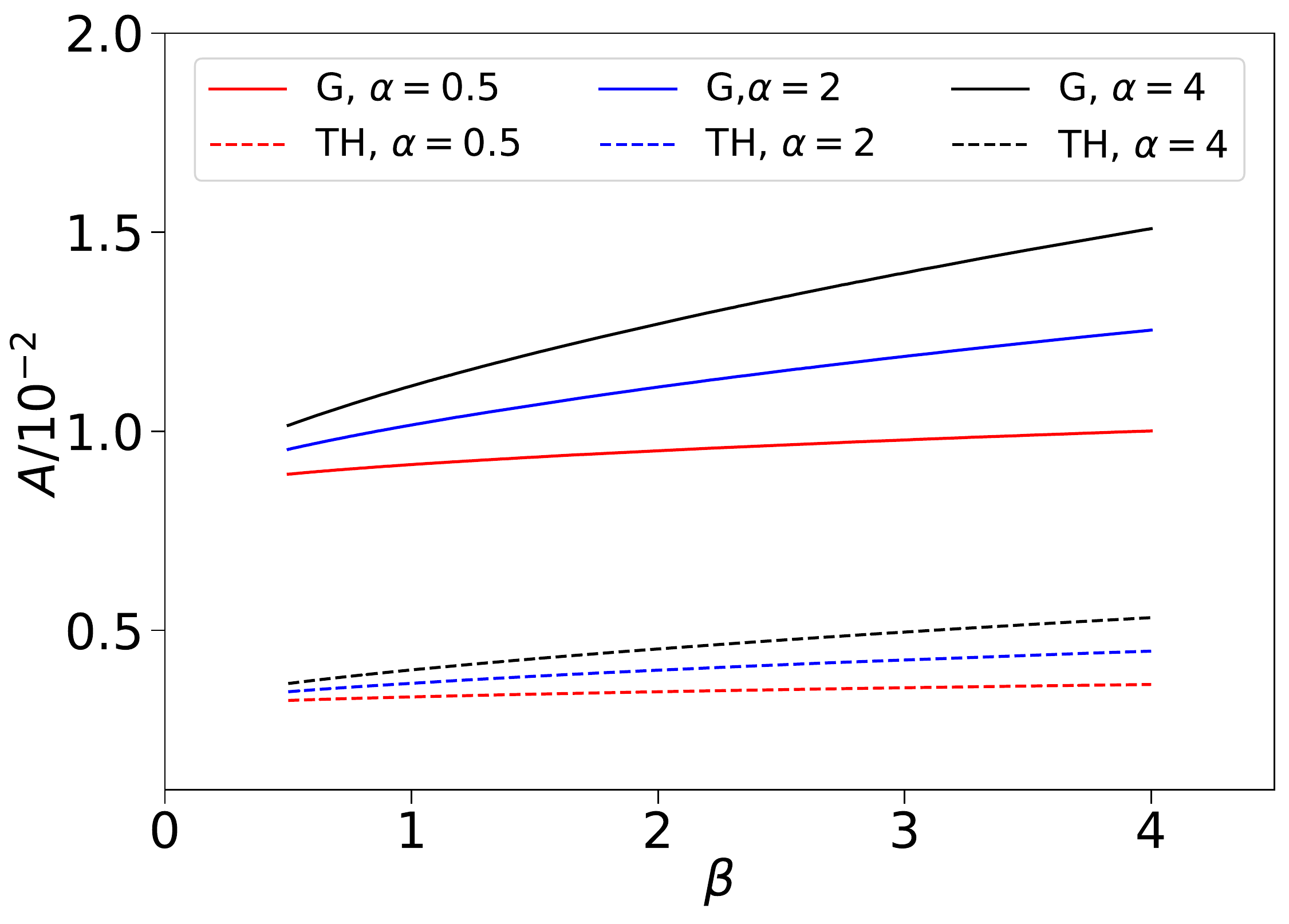}
	\caption{The required of amplitude $A$ to explain all the dark matter. The solid and dashed lines  are the results with the Gaussian and top-hat  window functions, respectively.}\label{fig:pl_all}
\end{figure}

Due to the failure of direct detection of particle dark matter, it is warranted to consider the possibility of PBHs as a DM candidate.  There are  no constraints on PBHs DM abundance at the two mass windows  $10^{-17}-10^{-15} M_{\odot}$ and $10^{-14}-10^{-12}M_{\odot}$,  as displayed in the left panel of figure \ref{fig:pl_allpbh_con}; therefore, PBHs with masses locating at these two windows can  make up all the dark matter. 
Combining  the parameterization \eqref{bpl:parametrize} and  PBHs mass function \eqref{app:fpbh:beta}, we can obtain the constraints on the primordial power spectrum if the PBHs DM make up all the dark matter, and the results are shown in figure \ref{fig:pl_all},  where  the peak scale  $k_p$ in parameterization \eqref{bpl:parametrize} is chosen as  $k_p=10^{13} ~\text{Mpc}^{-1}$.  
The solid lines denote the results with the Gaussian window function, and the dashed lines represent those with the real-space top-hat window function. 
The required amplitude $A$ of the primordial power spectrum is increasing  along with the parameter $\alpha$ and $\beta$. 
The reason is that the fraction of PBHs in the dark matter is the integration of mass function among the whole range as shown in equation \eqref{fpbh:tot}, so a narrower peak in the primordial power spectrum requires a larger amplitude $A$ to give the same fraction. A narrower peak is from a  pair of larger $\alpha$ and $\beta$,  so the amplitude $A$ increases along with the parameter $\alpha$ and $\beta$.  For the top-hat window  function,  the smallest value of the required amplitude  is $A\approx 3.24\times 10^{-3}$;  and for the Gaussian window function,  it is  $A\approx 8.92\times 10^{-3}$, which is about three times larger than that with the top-hat window function.
For the top-hat window  function,  the largest value of the required amplitude  is  $A\approx 5.32\times 10^{-3}$; and for the Gaussian window function,  it is $A\approx 1.51\times 10^{-2}$, which is also about three times larger than that with the top-hat window function.   
The largest value of $A$ is around $1.7$ times larger than the smallest value   for each window function.

\begin{figure}[htbp]
	\centering
	\includegraphics[width=0.95\columnwidth]{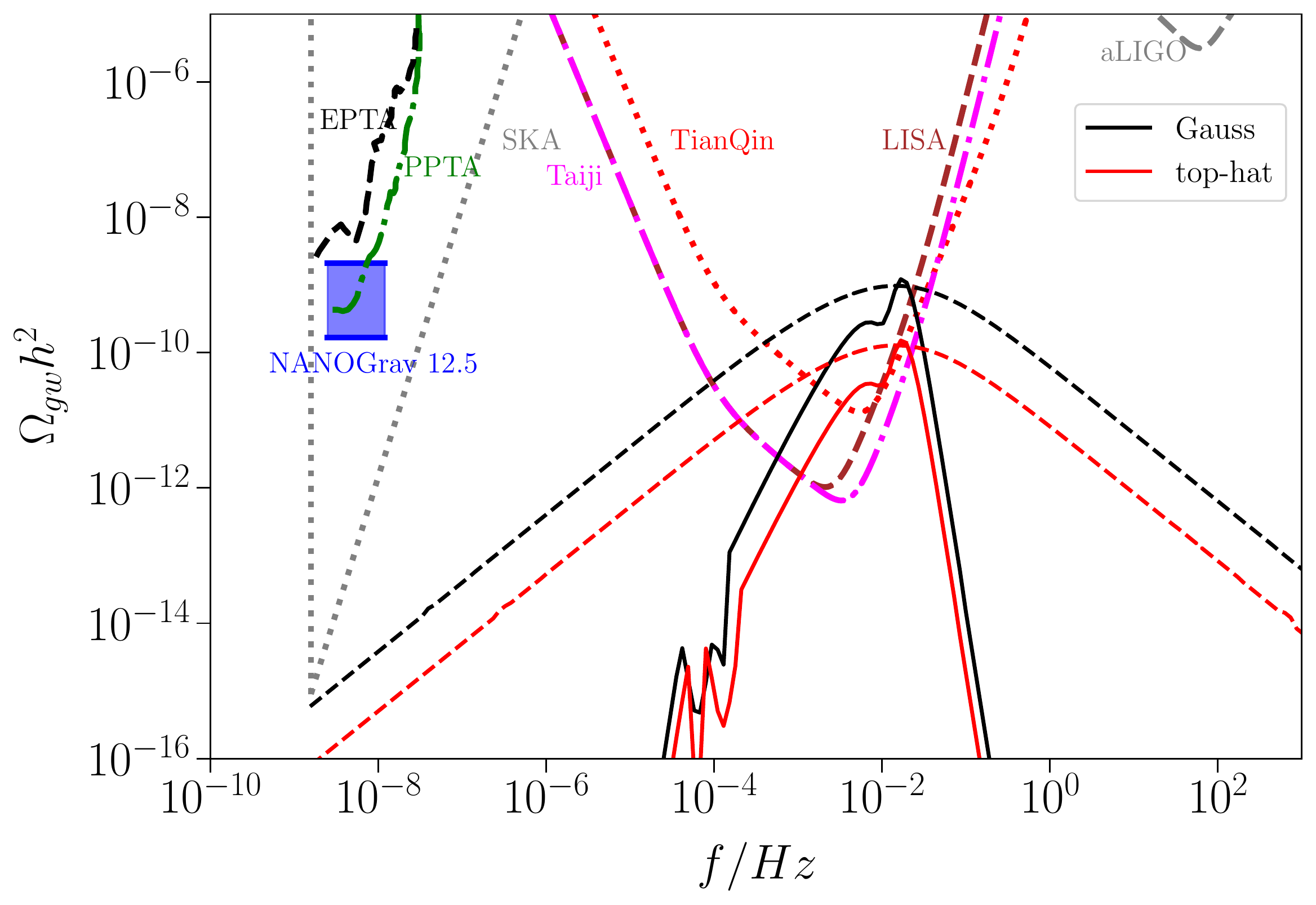}
	\caption{The  convex curves denote the energy density of  SIGWs, where the corresponding PBHs can explain all dark matter. 
		The black and  red curves  represent the situations where the Gaussian window function and  real-space  top hat window function are chosen to calculate the corresponding PBH mass function, respectively.
		The dashed and solid lines are from the parameterization \eqref{bpl:parametrize} with  the broadest peak and the narrowest peak, respectively.  }\label{fig:pl_all:gw}
\end{figure}

For the case where PBHs explain all the dark matter, the corresponding  SIGWs are displayed in figure \ref{fig:pl_all:gw}. The black lines denote the energy density of SIGWs where the Gaussian window function is chosen to calculate the corresponding PBH mass function, and the red lines  denote those with the top-hat window function being chosen.  
The dashed lines are the energy density of the GWs induced from the primordial
power spectra with the broadest peak, $\alpha=\beta=0.5$; and the solid lines are from the primordial power spectra with the narrowest peak,  $\alpha=\beta=4$.
If the PBHs formed from the primordial curvature perturbations can explain all the dark matter, the corresponding SIGWs will be detected by the future space-based GW detectors, and the profile of SIGWs can determine the profile of the peak in the primordial curvature
power spectrum. If the future space-based GW detectors do not detect the SIGWs, it indicates that the PBHs can only account for a part of dark matter.

\subsection{Constraints from  SIGWs}
The scalar-induced gravitational waves generated during the radiation domination are the components of the stochastic background of gravitational waves. The North American Nanohertz Observatory for Gravitational Wave (NANOGrav) Collaboration
has  published an analysis of the $12.5$yrs pulsar timing array (PTA) data, where strong evidence of a stochastic process with a common amplitude and a common spectral slope across pulsars was found  \cite{NANOGrav:2020bcs}. The same signal is also detected by   other pulsar
timing array groups \cite{Goncharov:2021oub,Antoniadis:2022pcn}. 
Although this process lacks quadrupolar spatial correlations,  it is worth interpreting as a stochastic GW signal, which the SIGWs can explain with frequencies around  $10^{-9}$ Hz \cite{DeLuca:2020agl,Inomata:2020xad, Vaskonen:2020lbd, Domenech:2020ers, Yi:2021lxc}. 

\begin{figure*}[htbp]
	\centering
	\includegraphics[width=1.7\columnwidth]{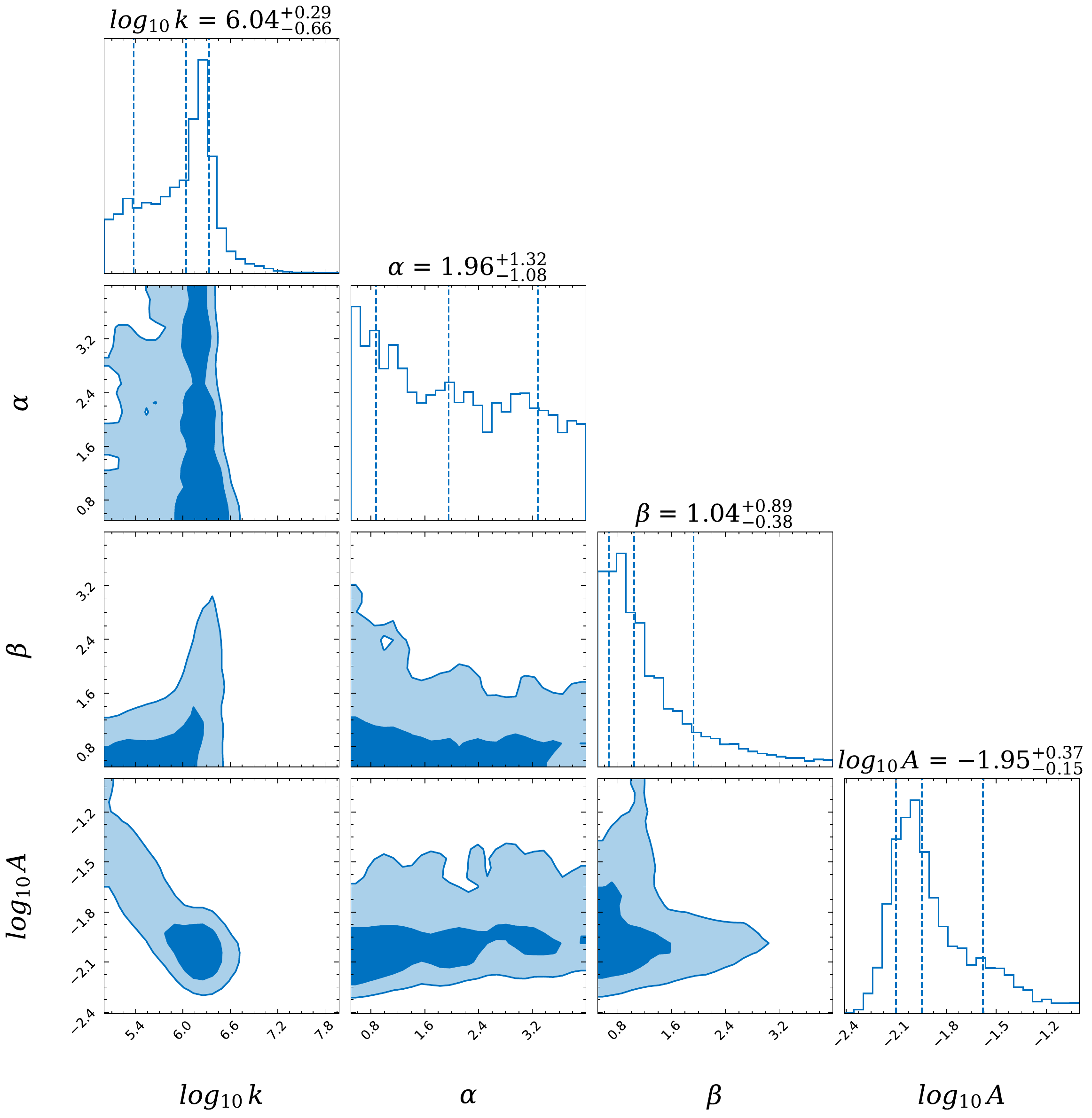}
	\caption{The posteriors on the parameters in  parameterization   \eqref{bpl:parametrize} from the first five   frequency bins of  NANOGrav 12.5yrs data set,
		and the shaded areas denote $1\,\sigma$ and $2\,\sigma$ confidence regions.}\label{fig:nanosigw}
\end{figure*}

This paper constrains the power spectrum of the primordial curvature perturbation from the  NANOGrav 12.5yrs data, assuming that    the NANOGrav 12.5yrs signals are from SIGWs.  
We follow the analysis  in ref. \cite{Moore:2021ibq} and focus on  the results of the NANOGrav free-spectrum analysis, where the signal in each frequency bin is fitted separately.  Only the posteriors on the first five frequency bins  are used due to the most constraining measurements coming at the lowest frequencies \cite{NANOGrav:2020bcs,Moore:2021ibq}.  
The public data products are from \href{https://data.nanograv.org/}{https://data.nanograv.org}.
The log-likelihood function is obtained by evaluating the  energy density of the SIGWs  at the five values $f_i$ and summing the log probability density functions of the five independent kernel density estimates at these values \cite{Moore:2021ibq}.  
%Sampling was performed with the DYNESTY\cite{dynesty}  implementation of the nested sampling algorithm \cite{NestedSampling}. 
The posteriors on the parameters of parameterization \eqref{bpl:parametrize}  are shown in  figure \ref{fig:nanosigw}, where the sampling is performed with the DYNESTY \cite{Speagle:2019ivv}   implementation of the nested sampling algorithm  \cite{NestedSampling}; and the marginalized posterior distributions lead to the following mean values and one-sigma confidence intervals,
\begin{gather}\label{beta:value1}
	\log_{10}A=-1.95_{-0.15}^{+0.37},\quad \log_{10}k=6.04_{-0.66}^{+0.29},
\end{gather} 
\begin{gather}\label{beta:value2}
	\alpha=1.96_{-1.08}^{+1.32},\quad \beta=1.04_{-0.38}^{+0.89}.
\end{gather} 
By choosing  these best parameter values \eqref{beta:value1} and \eqref{beta:value2}, the power spectrum with the  broken-power-law-parameterization \eqref{bpl:parametrize} is  displayed in the right panel of figure \ref{fig:pl_allpbh_con} and denoted by the black line. The corresponding 
mass function of PBHs calculated with peak theory and the Gaussian window function are shown in the left panel of figure \ref{fig:pl_allpbh_con} and represented by the black line, which are under the PBHs observational constraints. For the peak  theory with the top-hat window function, the mass function  at the peak is about $f_\text{PBH}(M_\text{PBH}^\text{peak})\sim 10^{4}$, exceeding the PBHs observational constraints, which can also be understood from  the right panel of figure \ref{fig:pl_allpbh_con}, so  we don't  display them in the left panel of  figure \ref{fig:pl_allpbh_con}.   

\begin{figure*}[htbp]
	\centering
	\includegraphics[width=1.2\columnwidth]{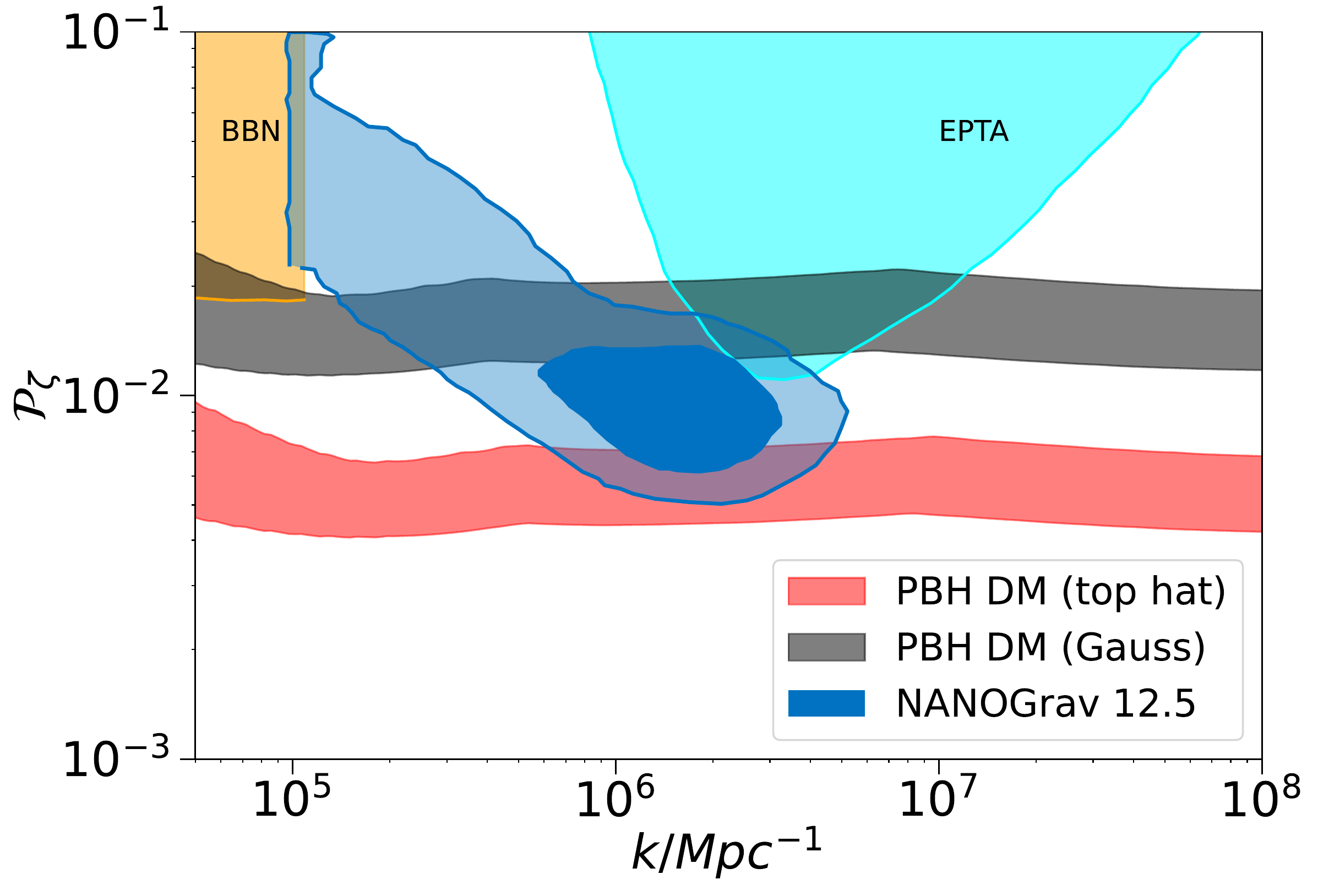}
	\caption{The constraints on the primordial curvature power spectrum. The orange and cyan  regions are excluded by BBN \cite{Inomata:2016uip}  and EPTA \cite{Inomata:2018epa}. The blue  area is  the allowed region  to explain the NANOGrav 12.5 yrs data sets; the two contours denote the $1\sigma$ and $2\sigma$ confidence regions, respectively. The black and red bands are the upper limits constrained from the PBHs observational data, as displayed in the right panel of figure \ref{fig:pl_allpbh_con}.}\label{fig:nanoak}
\end{figure*}

In figure \ref{fig:nanoak}, we take together the posteriors distribution from NANOGrav 12.5yrs data and other constraints on the primordial power spectrum.
The blue regions are the $1\,\sigma$ and  $2\,\sigma$ posteriors on the amplitude and scale, respectively.  The orange and cyan regions are excluded by the BBN  \cite{Inomata:2016uip} and EPTA  \cite{Inomata:2018epa}. The black and red bands are the  upper limits constrained  from the PBHs observational data, as displayed in the right panel of figure \ref{fig:pl_allpbh_con}.

\section{Conclusion}
The primordial black holes and scalar-induced gravitational waves  can be produced from the large scalar perturbations in the early Universe. They can tell us  information about the small-scale primordial curvature perturbation generated in the inflation. 
To form enough PBHs DM and induce detectable GWs, the amplitude of the primordial curvature power spectrum should be around  $\mathcal{P}_\zeta\sim \mathcal{O}(0.01)$, 
which is about seven orders of magnitude larger than the constraints from CMB observational data at large scales.  
Generally, the enhanced region of the primordial power spectrum can be parameterized as the power law form. 
For the whole primordial power spectrum, we use the  broken power law form to parameterize their profile, where the parameterization form is  $\mathcal{P}_\zeta(k)=A(\alpha+\beta)/[\beta(k/k_p)^{-\alpha}+\alpha(k/k_p)^\beta]+A_*(k/k_*)^{n_{s*}-1}$ with $k_*=0.05\text{Mpc}^{-1}$, $n_{s*}=0.965$, and $A_*=2.1\times 10^{-9}$.
The  primordial  curvature power spectrum with the  broken power form can be produced from many inflation models, such as nonminimal derivative coupling inflation,  scalar-tensor inflation, Gauss-Bonnet inflation, and K/G inflation.

The fraction of PBHs  in the dark matter is calculated by the peak theory, where both  Gaussian and top-hat window functions are considered. 
With the help of the fraction equation, we obtain  the constraints on the primordial curvature power spectrum from the present primary PBHs DM observational data. The results are displayed in  the right panel of figure \ref{fig:pl_allpbh_con}. 
For the choice of the top-hat window function, the constraints on  the primordial curvature power spectrum from PBHs DM are more robust than those from BBN and EPTA. For the Gaussian window function, the constraints from PBHs DM are weaker than those from   EPTA.  
The uncertainty of the constraints caused by the choice of the window function is more significant than that by the profile of the peak in the primordial curvature power spectrum; we should take care of the choice of the window function.  
Under the constraints of  PBHs DM,
the corresponding SIGWs  with frequencies around  $10^{-9}$ Hz,  $10^{-3}$ Hz, and $10^2$ Hz are shown in   figure \ref{fig:pl_allpbh_con:gw}.
The SIGWs with frequencies around  $10^{-9}$ Hz and  $10^{-3}$ Hz can be detected by PTA and the future space-based detectors, respectively. 
The SIGWs with frequencies around $10^2$ Hz cannot be detected  presently, and we will get little information about the early Universe from them. 
The  required amplitudes of the primordial curvature power spectrum to explain all the dark matter are displayed in figure \ref{fig:pl_all}, and they depend on the shape of the spectrum  peak  and window function. 
For the top hat window function, to explain all the dark matter, the amplitude of the primordial power spectrum  with the narrowest peak requires $A\approx 5.32\times 10^{-3}$  and $A\approx 3.24\times 10^{-2}$ with the broadest peak; for the Gaussian window function, it requires $A\approx 1.51 \times 10^{-2}$ with the narrowest peak and $A\approx 8.92\times 10^{-3}$  with the broadest peak. 
The largest required  amplitude is about $1.7$ times larger than the smallest for each window function. The corresponding SIGWs can be detected by future space-based detectors, and the SIGWs can be used to verify whether PBHs can explain all the dark matter.

The constraints on the primordial curvature power spectrum from  NANOGrav 12.5yrs data sets are displayed in  figure \ref{fig:nanosigw} by assuming the NANOGrav signals are from SIGWs. 
The mean values and one-sigma confidence intervals of the broken power law parameterization of primordial curvature power spectrum obtained from the first five frequency bins of  NANOGrav 12.5yrs data sets are   $\log_{10}A=-1.95_{-0.15}^{+0.37}$, $\log_{10}k=6.04_{-0.66}^{+0.29}$,  $\alpha=1.96_{-1.08}^{+1.32}$,  and $\beta=1.04_{-0.38}^{+0.89}$.  The allowed region of the curvature power spectrum from the NANOGrav 12.5yrs data sets, the upper limits  from the PBH observational data,  and the excluded regions from BBN and EPTA are taken together in figure  \ref{fig:nanoak}. 

In conclusion, we give the constraints on the primordial curvature perturbations from the present primary  PBHs observational data,  where the fraction of PBHs in dark matter is calculated by the peak theory and the primordial curvature spectrum is parameterized by the broken power law form. 
The region of the primordial curvature spectrum to explain the NANOGrav 12.5yrs data sets is also obtained by assuming 
the NANOGrav signals are from SIGWs.

\begin{acknowledgement}
We thank Xing-Jiang Zhu, Zu-Cheng Chen, Xiao-Jin Liu,  Zhi-Qiang You, and  Shen-Shi Du for useful discussions.
This research is supported  by  the National Natural Science Foundation
of China under Grant No. 12205015 and the supporting fund for young researcher of Beijing Normal University under Grant No. 28719/310432102.
\end{acknowledgement}

%\bibliographystyle{spphys}
%\bibliography{main}

\end{document}